\documentclass[showpacs,preprint,showkeys,superscriptaddress,amssymb,
               nofootinbib,aps]{revtex4}

\makeatletter
\renewcommand{\p@subsection}{}
\renewcommand{\p@subsubsection}{}
\makeatother
\usepackage{array,multirow}
\usepackage{mdwlist}
\usepackage{amsmath}
\usepackage{amsthm}

\usepackage{amssymb}
\usepackage{bm}
\usepackage{blindtext}
\usepackage{enumitem}
\usepackage{natbib}
\usepackage{graphicx}

\def\be{\begin{equation}}
\def\ee{\end{equation}}
\def\ba{\begin{eqnarray}}
\def\ea{\end{eqnarray}}

\begin{document}
\title{\Large Charged dust solutions for the warp drive spacetime}

\author{Osvaldo L. Santos-Pereira}\email{olsp@if.ufrj.br}
\affiliation{Physics Institute, Universidade Federal do Rio de Janeiro,
        Brazil}
\author{Everton M.\ C.\ Abreu}\email{evertonabreu@ufrrj.br}
\affiliation{Department of Physics, Universidade Federal Rural do Rio
       de Janeiro, Serop\'edica, Brazil}
\affiliation{Department of Physics, Universidade Federal de Juiz de
       Fora, Brazil}
\affiliation{Applied Physics Graduate Program, Physics Institute,
       Universidade Federal do Rio de Janeiro, Brazil}
\author{Marcelo B.\ Ribeiro}\email{mbr@if.ufrj.br}
\affiliation{Physics Institute, Universidade Federal do Rio de
       Janeiro, Brazil}
\affiliation{Applied Physics Graduate Program, Physics Institute,
       Universidade Federal do Rio de Janeiro, Brazil}
\affiliation{Valongo Observatory, Universidade Federal do Rio de
       Janeiro, Brazil}
\date{\today}
 
\begin{abstract}
\noindent  The Alcubierre warp drive metric is a spacetime construction
where a massive particle located inside a spacetime distortion, called
warp bubble, travels at velocities arbitrarily higher than the velocity
of light. This theoretically constructed spacetime geometry is a
consequence of general relativity where global superluminal velocities,
also known as warp speeds, are possible, whereas local speeds are
limited to subluminal ones as required by special relativity. In this
work we analyze the solutions of the Einstein equations having charged
dust energy-momentum tensor as source for warp velocities. The Einstein
equations with the cosmological constant are written and all solutions
having energy-momentum tensor components for electromagnetic fields
generated by charged dust are presented, as well as the respective energy
conditions. The results show an interplay between the energy conditions
and the electromagnetic field such that in some cases the former can be
satisfied by both positive and negative matter density. In other cases
the dominant and null energy conditions are violated. A result connecting
the electric energy density with the cosmological constant is also
presented, as well as the effects of the electromagnetic field on the
bubble dynamics.
\end{abstract}

\maketitle

%%%%%%%%%%%%%%%%%%%%%%%%%%%%%%%%%%%%%%%%%%%%%%%%%%%%%%%%%%%%%%%%%%%%%%
\section{Introduction}
\renewcommand{\theequation}{1.\arabic{equation}}
\setcounter{equation}{0}

It has been known for some time that general relativity allows for
particles to travel globally with superluminal velocities because
special relativity limits the particles velocities to subluminal ones
only locally. In order words, special relativity basically states that
the light speed limit must be obeyed inside a local light cone. The
\textit{warp drive metric} proposed by Alcubierre \cite{Alcubierre1994}
satisfies both conditions, since it advances a geometrical construction
characterized by a spacetime distortion, called \textit{warp bubble},
such that a particle inside this bubble travels with superluminal
velocity in global terms, whereas locally its speed remains subluminal.
more specifically, locally the warp bubble guarantees that the
particle's velocity is kept below the light speed, whereas outside the
local light cone, created by the bubble distortion, the whole bubble
structure travels with superluminal velocities, or \textit{warp speeds.}
The warp bubble is created in such a way that the spacetime in front of
it is contracted while behind the bubble the spacetime is expanded.
In its original proposition and first studies it was foresaw that the
warp metric would violate the energy conditions, as well as supposedly
requiring huge quantities of negative energy density.

After Ref.\ \cite{Alcubierre1994} several papers tried to comprehend
the main aspects of the warp drive metric. For example, Ref.\
\cite{FordRoman1996} discussed some quantum inequalities that should
be valid as a result of the \textit{Alcubierre warp drive metric},
concluding that large amounts of negative energy would be needed to
convey particles with small masses across small distances at warp
speeds. Therefore, they figured that prohibitive huge quantities of
negative energy density would be necessary to generate a warp bubble.
Also dealing with quantum inequalities, Ref.\ \cite{Pfenning1997} 
computed the limits required for the energy values and the bubble
parameters necessary for the existence of the warp drive. The
conclusion reached by this author was that the energy needed for a
warp bubble is ten orders of magnitude greater than the total mass of
all the visible universe, also negative.

Looking at the same problem of the physics of superluminal propulsion
systems for interstellar travel, but from a different viewpoint,
Krasnikov \cite{Krasnikov1998} analyzed the scenario of a massive
particle making a round trip between two distant points in space at
speeds faster than a photon. He questioned that this is not viable
when reasonable conditions for global hyperbolic spacetimes are made.
He analyzed in details some peculiar spacetime topologies, supposing
that, for some of them, they need tachyons for superluminal trips to
occur. He also assumed the need for a possible particular spacetime
prearranged with some devices along the travel path, which would be
set up and activated as needed in order to the superluminal travel to
occur without tachyons. Such spacetime constructions was called as
the \textit{Krasnikov tube} by Ref.\ \cite{EveretRoman1997}. 

The metric proposed by Krasnikov was further generalized by Everett
and Roman \cite{EveretRoman1997} by conceiving a tube along the path
of the particle connecting Earth to a distant star. Inside this tube
the spacetime is flat, however the lightcones are opened out in such
a way as to allow the superluminal travel in one direction. One of
the issues analyzed in Ref.\ \cite{EveretRoman1997} is that since the
Krasnikov tube does not involve closed timelike curves we are able to
construct a two way non-overlapping tube system such that it would
work as a time machine. It was also demonstrated in Ref.\
\cite{EveretRoman1997} that a great quantity of negative energy
density is necessary for the Krasnikov tube to function. These
authors also used the generalized Krasnikov tube metric to compute an
energy-momentum tensor (EMT) which would be positive in some particular 
regions. 

Further studies concerning the metric proposed by Everett and Roman
\cite{EveretRoman1997} were carried out by Lobo and Crawford
\cite{Lobo2003,Lobo2002}. They investigated in detail the metric and
the respective EMT obtained from it and if it were possible for a
superluminal travel to exist without violating the weak energy
condition. Quantum inequalities used in Ref.\ \cite{EveretRoman1997} 
were also explored.

Van de Broeck \cite{Broeck1999} demonstrated that a minor
modification of the Alcubierre geometry would reduce the total energy
necessary for the creation of the warp bubble. By introducing a
modification of the original warp drive metric the total negative
mass-energy necessary to describe the spacetime distortion capable of
warp speeds would be of the order of some solar masses. Natario
\cite{Natario2002} questioned if both the expansion and contraction
of the space of the bubble is a matter of choice. He suggested a new
version of the warp drive metric with zero expansion. Lobo and Visser
\cite{LoboVisser2004} discussed that for the Alcubierre warp drive,
and its version proposed in Ref.\ \cite{Natario2002}, the center of
the bubble must be massless. They presented a linearized theory for
both concepts and found out that even for low speeds the negative
energy stored in the warp fields must be just a relevant part of the
mass of the particle at the center of the warp bubble. White
\cite{White2003,White2011} depicted how a warp field interferometer
could be implemented at the Advanced Propulsion Physics Laboratory. 
Lee and Cleaver \cite{cleaver1,cleaver2} looked at how external
radiation might affect the Alcubierre warp bubble, possibly making
it energetically unsustainable, and how a proposed warp field
interferometer could not detect spacetime distortions. Mattingly
et al.\ \cite{cleaver3} discussed curvature invariants in the
Natario warp drive.

One aspect that is often overlooked regarding the Alcubierre warp
drive metric is that the original proposal did not come from
solving the Einstein field equations (Alcubierre 2018, private
communication), but as a geometrical construction theoretically
capable of generating warp speeds. Indeed, the original proposal
of the warp drive metric was not accompanied by a dynamical equation,
which is the case when a metric is reached from solutions of the
Einstein equations.

In a previous paper \cite{nosso-1} we started from this realization
and then endeavored to actually solve the Einstein equations using
the simplest possible mass-energy distribution, incoherent matter or
dust, as a starter in order to verify if this distribution were
actually capable of creating a superluminal warp field, that is, a
warp bubble. Although the results went back to vacuum, the solutions
have indeed generated a dynamical equation for the warp metric
regulating function $\beta$ (see below), which was found to obey in
a particular case the Burgers equation for inviscid fluid with
shockwaves in the form of plane waves \cite[see also Ref.][]{nosso-2}. 

In this work we intend to pursue a similar path, that is, to discuss
solutions of the Einstein equations having the Alcubierre warp
drive geometry and considering a non null cosmological constant. To
do that we considered a charged dust capable of generating an
electromagnetic field, wrote the EMT for both components, solved the
equations and wrote the respective energy conditions. The results
showed an interplay between the energy conditions and the
electromagnetic field such that in some cases the former can be
satisfied by both positive and negative matter density. In other cases
the dominant and null energy conditions are violated. A result
connecting the electric energy density with the cosmological constant
is also presented, as well as the effects of the electromagnetic field
on the bubble dynamics.

The plan of the paper is as follows. Section 2 presents both a brief
review of the warp drive metric and electromagnetism in curved
spacetimes necessary for a self contained paper. In Section 3 we
calculated the electromagnetic EMT and the respective Einstein tensor
components. In Section 4 we analyzed the energy conditions and Section
5 is dedicated to the investigation of conditions concerning the
electric and magnetic fields under the warp drive metric. Section 6
presents our conclusions and final remarks. 

%%%%%%%%%%%%%%%%%%%%%%%%%%%%%%%%%%%%%%%%%%%%%%%%%%%%%%%%%%%%%%%%%%%%%%
\section{Basic concepts}
\renewcommand{\theequation}{2.\arabic{equation}}
\setcounter{equation}{0}

\subsection{Warp drive metric}

From special relativity it is well known that nothing moves faster
than light, but in the scope of general relativity, the spacetime 
is not flat anymore, but dynamic. This allows the possibility of 
exotic solutions such as wormholes and warp drive, for example. 
This last one is a way of going from a point A to a point B in space 
in times arbitrarily smaller than the light would take to travel 
between those points.

In \cite{Alcubierre1994} it was described a possible way that a
mass particle could travel from one point to another in spacetime
measured by an external observer in a time interval smaller than the
light would travel the same distance. In few words, consider a
particle that leaves an inertial reference frame $A$, which remains
at rest, towards another inertial reference frame, also at rest,
named $B$, at a distance $D$ from $A$. The particle is inside a
bubble that can modify the spacetime in a way that the space behind
the bubble is expanded whereas it is contracted in front of if.
This dynamics allows the particle inside the bubble to travel the
distance $D$ in a time less than $D/c$, where $c$ is the speed of
light, as measured by external observers distant from the bubble,
although it is still moving inside a light cone, which means that
the particle does not travel faster than light locally.

The warp drive metric \cite{Alcubierre1994}, using Cartesian
coordinates $x^\mu = (t,x,y,z)$, is given by
\be
ds^2 = - (1 - \beta^2) dt^2 - 2 \beta(r_s,t) dx \, dt + 
dx^2 + dy^2 + dz^2\,\,,
\label{wdmetric}
\ee
where the term $\beta(r_s,t)$ is the shift vector, a boost in the 
$x$-direction where the particle describes its trajectory inside the 
bubble. The shift vector is given by
\be
\beta(r_s,t) = v_s(t) f(r_s)\,\,,
\ee
where $v_s(t) = dx_s/dt$ is the warp bubble speed, $x_s$ is
the coordinate for the center of the bubble, $f = f(r_s)$ 
is the regulating function \cite{Alcubierre1994}, which 
regulates the warp bubble shape. The parameter $r_s$ is the 
radius of the bubble given by
\be
r_s = \sqrt{[x - x_s(t)]^2 + y^2 + z^2}\,\,.	
\ee
In this work we shall assume $\mu_{{}_0} = \epsilon_{{}_0} = 1$
and $G = c = 1$.

%%%%%%%%%%%%%%%%%%%%%%%%%%%%%%%%%%%%%%%%%%%%%%%%%%%%%%%%%%%%%%%%%%%%
\subsection{Electromagnetism in curved spacetime}

In this section we will calculate the EMT for this warp drive metric.
But first we will introduce the electromagnetic energy-momentum tensor
(EEMT) $T_{\alpha \beta}$ using, of course, the electromagnetic field 
strength tensor $F_{\alpha\beta}$.

It is well known that Maxwell electromagnetism is consistent
with special relativity, the Lorentz force law, that the Maxwell
equations are valid for any inertial reference system, and that they  
can be put in what is known as the covariant formulation, which
means to describe electromagnetism in special relativity language
in a manifestly invariant form under Lorentz transformations. This
formalism is constructed in a flat spacetime with Minkowski metric
in Cartesian coordinates given by
\be
ds^2 = - dt^2 + dx^2 + dy^2 + dz^2\,\,.
\label{Mink}
\ee

Due to the manifest covariance of the Maxwell equations 
in spacetime notation, if partial derivatives are replaced
by covariant derivatives the extra terms cancels out and 
the equations remain the same, making it possible to 
substitute the Minkowski metric $\eta_{\mu\nu}$ for a 
curved spacetime metric $g_{\mu\nu}$ in a general 
curvilinear coordinate system.

The 4-gradient is given by the following expression
\be
\partial^\mu = \frac{\partial}{\partial x^\mu} = 
\left(\frac{\partial}{\partial t}, - \nabla \right) \,\,,
\ee
and we have that $g_{\alpha \beta}$, and the 
4-gradient in covariant form is $\partial_\alpha = 
g_{\alpha\beta} \partial^\beta$. 

The electric $\mathbf{E}$ and magnetic $\mathbf{B}$ fields
are described through the electromagnetic 4-potential, which 
is a covariant 4-vector with the electric scalar potential 
$\phi$ as its first component and the magnetic vector 
potential $\mathbf{A}$ as the other components,
\be
A^{\alpha }=\left(\phi,{\mathbf{A}}\right)\,\,.
\ee
The electric and magnetic fields above can be written in tensor 
notation, such as
\begin{align}
B^a &= \epsilon^{abc}\Big(\partial_b A_c - \partial_c A_b \Big)\,,
\\
E^a &= \partial_{{}_0} A_a - \partial_a A_{{}_0}\,\,.
\end{align}

In matrix form, the field strength $F^{\alpha \beta}$ can be 
written as,
\be
F^{\alpha \beta} = 
\begin{pmatrix}
0   & -E_1 & -E_2  & -E_3 \\
E_1 & 0    & -B_3  & B_2 \\
E_2 & B_3  & 0     & -B_1 \\
E_3 & -B_2 & B_1   & 0
\end{pmatrix} \,\,,
\label{contraf1}
\ee
where the electric field $\mathbf{E}$ and the magnetic field  
$\mathbf{B}$ are in vector form
\begin{align}
\mathbf{E} &= (E_1, E_2, E_3),
\\
\mathbf{B} &=	(B_1, B_2, B_3),
\end{align}
The covariant form of the electromagnetic tensor in general
curved spacetime is given by the following expression,
\be
F_{\alpha \beta} = g_{\alpha \mu} \,g_{\beta \nu}\, F^{\mu \nu} \,.
\label{covf1}
\ee

The EMT for curved space time can be written 
as, where we use here that the signature is $(-+++)$, 
\be
T_{\alpha\beta} = \frac{1}{4\pi}\left(
\frac{1}{4} g_{\alpha\beta}F_{\gamma\nu}F^{\gamma\nu}
- g^{\gamma\nu}F_{\alpha\gamma}
F_{\beta \nu}\right).
\ee

If an extra term ($\mu\, u_{{}_\alpha} u_{{}_\beta}$) is added to 
EEMT it would describe a charged matter with density $\mu$ and 
proper velocity $u_\alpha$ \cite{dInverno1992}.   The EMT can be 
given by 
\begin{align}
\nonumber T^{\alpha\beta} &= T^{\text{(dust)}} + T^{\text{(elec)}} \\
&= \frac{\mu}{4\pi} u^\alpha u^\beta  + 
\frac{1}{4\pi}\left(\frac{1}{4}g^{\alpha\beta}F_{\nu\sigma}F^{\nu\sigma}
- F^{\alpha\nu}F^\beta{}_\nu\right)\,\,,
\label{EEMT1}
\end{align}
which is the EEMT for the dust embedded in an electromagnetic field in
a curved spacetime. It is very clear in the last equation what each
part of the tensor means.

%%%%%%%%%%%%%%%%%%%%%%%%%%%%%%%%%%%%%%%%%%%%%%%%%%%%%%%%%%%%%%%%%%%%
\section{The energy-momentum and Einstein tensor components}
\renewcommand{\theequation}{3.\arabic{equation}}
\setcounter{equation}{0}

From now on we will discuss the conditions for the energy density 
$T_{00}$ from the EMT to be positive and how the radiant matter 
density $\mu$ together with the electromagnetic components 
influence the warp drive and how higher orders of the warp drive 
shift $O(\beta^2)$ influence the spacetime. We will discuss 
interesting expressions found from the computation of Einstein equations 
that provide us some insight on the plausibility of the warp drive concept.

%%%%%%%%%%%%%%%%%%%%%%%%%%%%%%%%%%%%%%%%%%%%%%%%%%%%%%%%%%%%%%%%%%%%
\subsection{Electromagnetic energy momentum tensor}

Considering the 4-velocity of the Eulerian observers $u_{\alpha} = 
(-1,0,0,0)$ \cite{Alcubierre1994}, the dust matter radiation 
part $\mu u_\alpha u_\beta$ from Eq. \eqref{EEMT1} has only one 
non zero component, i.e.,  $T^{\text{(dust)}}_{00} = \mu$.    The non zero
and non redundant components of the EEMT are given by
\begin{align}
\nonumber 4\pi T_{00} &= \mu + \frac{1}{2}(B^2 + E^2)
+ \left(B_3 E_2 - B_2 E_3\right)\beta 
+ \frac{\beta^2}{2}\left(B^2 - 2B_1^2 - E^2 - E_3^2\right)
\\
&
+ \left(B_2 E_3 - B_3 E_2\right) \beta^3 
+ \frac{\beta^4}{2}\left(E_2^2 + E_3^2\right),
\label{eemt00}
\end{align}

\be
4\pi T_{01} = B_2 E_3 - B_3 E_2 
+ \frac{\beta}{2}\left(2B_1^2 - B^2 + E^2\right)
+ \left(B_3 E_2 - B_2 E_3\right) \beta^2   
-\frac{\beta^3}{2}\left(E_2^2 + E_3^2\right)
\label{eemt01}
\ee

\be
4\pi T_{02} = B_1 E_3 \beta^2 + B_1 B_2 \beta + B_3 E_1 - B_1 E_3, 
\label{eemt02}
\ee

\be
4 \pi T_{03} = - B_1 E_2 \beta^2 + B_1 B_3 \beta - B_2 E_1 + B_1 E_2, 
\label{eemt03}
\ee

\be
4\pi T_{11} = \frac{\beta^2}{2} \left(E_2^2 + E_3^2\right) + 
\frac{1}{2}\left(B^2 - 2B_1^2 + E^2 - 2E_1^2\right) 
- \left(B_3 E_2 - B_2 E_3\right) \beta, 
\label{eemt11}
\ee

\be
4\pi T_{12} = -B_1 E_3 \beta - B_1 B_2 - E_1 E_2, 
\label{eemt12}
\ee

\be
4\pi T_{13} = B_1 E_2 \beta - B_1 B_3 - E_1 E_3, 
\label{eemt13}
\ee

\be
4\pi T_{22} = \frac{\beta^2}{2}\left(E_2^2 - E_3^2\right)  - 
\left(B_3 E_2 + B_2 E_3\right) \beta +
\frac{1}{2} \left(B^2 - 2B_2^2 + E^2 - 2E_2^2\right), 
\label{eemt22} 
\ee

\begin{align}
4\pi T_{23} &= E_2 E_3 \beta^2 + 
\left(B_2 E_2 - B_3 E_3\right) \beta - 
(B_2 B_3 + E_2 E_3) , 
\label{eemt23} 
\end{align}

\be
4\pi T_{33} = -\frac{1}{2}\left(E_2^2 - E_3^2\right) \beta^2 + 
\left(B_3 E_2 + B_2 E_3\right) \beta + 
\frac{1}{2}\left(B^2 - 2B_3^2 + E^2- 2E_3^2\right)\,\,.
\label{eemt33}
\ee

If we take $\beta \to 0$ in  the warp drive metric in Eq.\ \eqref{EEMT1},
it becomes asymptotically the Minkowski metric written in Eq.\
\eqref{Mink} and there is no warp drive. Using this result, the EEMT
will become the one calculated for the Minkowski metric
\be
T_{\alpha \nu} =
\begin{pmatrix}
4\pi \mu + \frac{1}{2}(E^2 + B^2) & -S_1         & -S_2         & -S_3         \\
-S_1                              & -\sigma_{11} & -\sigma_{12} & -\sigma_{13} \\
-S_2                              & -\sigma_{21} & -\sigma_{22} & -\sigma_{23} \\
-S_3                              & -\sigma_{31} & -\sigma_{32} & -\sigma_{33}
\end{pmatrix}.
\label{emtmatrix}
\ee
where $S_1$, $S_2$ and $S_3$ are the vector components of the Poynting
vector, $\mathbf{S} =  \mathbf{E} \times \mathbf{B}/\mu _0$, and $\sigma_{ij}$ 
are the nine components of the Maxwell stress tensor defined by 
\be
\sigma _{ij} = \epsilon_0 E_iE_j+ \frac{1}{\mu_0} B_iB_j - 
\frac{1}{2}\left(\epsilon_0 E^2 + \frac{1}{\mu_0} B^2\right)\delta_{ij}\,\,,
\label{maxst1}
\ee
which will be rewritten as 
\be
\sigma _{ij} = E_i E_j + B_iB_j - \frac{1}{2}\Big(E^2 + B^2\Big)
\delta_{ij}\,\,,
\label{maxst2}
\ee
where $i,\,j = 1,2,3$. We can notice that considering a warp drive 
spacetime background, the EEMT components have extra $\beta$-correction
terms in comparison with Minkowski background. In this scenario 
we can consider that the warp drive can generate electromagnetic fields
or at least it reinforces an already existing one, as it can be seen from
Eqs. \eqref{eemt00} to \eqref{eemt33}. In particular, the component
$T_{00}$, which is the energy density, has fourth order $\beta$-corrections.

%%%%%%%%%%%%%%%%%%%%%%%%%%%%%%%%%%%%%%%%%%%%%%%%%%%%%%%%%%%%%%%%%%%%%

\subsection{Energy density $T_{00}$}

Analyzing the EEMT $T_{00}$ term, which is given by Eq. 
\eqref{eemt00}, it shows that this term is a $\beta$-fourth order degree 
polynomial that has the form
\be
T_{00}(\beta) = \omega_{(0)} + \omega_{(1)} \beta + 
\omega_{(2)} \beta^2  + \omega_{(3)} \beta^3 + \omega_{(4)} \beta^4 \,\,,
\label{t00energy}
\ee
where the coefficients $\omega_{(k)}$ are given by the expressions 
\be
\omega_{(0)} = \mu + \frac{1}{2}(B^2 + E^2)\,\,,
\label{t00omega0}
\ee

\be
\omega_{(1)} = B_3 E_2 - B_2 E_3\,\,,
\label{t00omega1}
\ee

\be
\omega_{(2)} = \frac{1}{2}\left(B^2 - 2B_1^2 - E^2 - E_3^2\right)\,\,,
\label{t00omega2}
\ee

\be
\omega_{(3)} = B_2 E_3 - B_3 E_2\,\,,
\label{t00omega3}
\ee

\be
\omega_{(4)} = \frac{1}{2}\left(E_2^2 + E_3^2\right)\,\,.
\label{t00omega4}
\ee

As we said above, for $\beta = 0$, the warp drive
metric becomes the Minkowski metric and the only non zero component
of the $T_{00}$ term would be $\omega_{{}_0}$ as expected from 
Eq.\,\eqref{emtmatrix}. Assuming a ``weak" warp drive, that considers
only shift vector $O(\beta)$ terms, the condition for $\beta$, where
$T_{00} \approx \omega_{(0)} + \omega_{(1)} \beta > 0$ is
\be
\beta \:\: < \:\: \frac{\mu}{B_2 E_3 - B_3 E_2}
+ \frac{1}{2}\frac{B^2 + E^2}{B_2 E_3 - B_3 E_2}\,\,,
\label{t00lowbeta}
\ee
which means that for this condition the shift vector has an upper bound
limit which can be very large if $B_2 E_3 \approx B_3 E_2$. Now
considering an usual warp drive, with higher orders of the shift 
vector $O(\beta^2)$, the conditions for $T_{00} > 0$ requires that
the terms $\omega_{(2)}$ and $\omega_{(3)}$ from Eqs. 
\eqref{t00omega2} and \eqref{t00omega3} are positive, and consequently 
we have the following conditions
\be
B_2^2 + B_3^2 \:\:>\:\: E_1^2 + E_2^2 + 2 E_3^2 + B_1^2\,\,.
\label{relfieldcomp2}
\ee

\be
B_2 E_3 - B_3 E_2 > 0\,\;,
\label{relfieldcomp3}
\ee
where we have neglected the term $\omega_{(4)}$ in Eq.\ \eqref{t00omega4},
since it is always positive and it is multiplied by $\beta^4$. 
Both Eqs.\ \eqref{relfieldcomp2} and \eqref{relfieldcomp3} show 
connections between the components of the electromagnetic field in a way that 
the energy density $T_{00}$ is positive for the warp drive spacetime in an 
electromagnetic background with a cosmological constant.

%%%%%%%%%%%%%%%%%%%%%%%%%%%%%%%%%%%%%%%%%%%%%%%%%%%%%%%%%%%%%%%%%%%%%

\subsection{Einstein tensor components}

Let us now calculate the Einstein tensor components added by the 
cosmological constant. Considering
\be
G_{\mu\nu} = R_{\mu\nu} - \frac{1}{2}g_{\mu\nu}R - 
\Lambda g_{\mu\nu}\;\;,
\ee
where $g_{\mu\nu}$ is the warp drive metric tensor given in Eq.\
\eqref{wdmetric} and $\Lambda$ is the cosmological constant. So,
we have that 
\be
G_{00} =  \Lambda(1-\beta^2) - \frac{1}{4} 
(1 + 3\beta^2)
\left[
\left(\frac{\partial \beta}{\partial y} \right)^2 +  
\left(\frac{\partial \beta}{\partial z} \right)^2 
\right] 
- \beta \left(\frac{\partial^2 \beta}{\partial y^2} + 
\frac{\partial^2 \beta}{\partial z^2}\right),
\label{et00}
\ee

\be
G_{01} =  \Lambda \beta + \frac{3}{4} 
\beta \left[
\left(\frac{\partial \beta}{\partial y}\right)^2 
+ \left(\frac{\partial \beta}{\partial z}\right)^2 
\right] 
+ \frac{1}{2}\left(
\frac{\partial^2 \beta}{\partial y^2} 
+ \frac{\partial^2 \beta}{\partial z^2}
\right),
\label{et01}
\ee

\be
G_{02} = - \frac{1}{2}
\frac{\partial^2 \beta}{\partial x \partial y} 
- \frac{\beta}{2} 
\left(2\frac{\partial \beta}{\partial y}
\, \frac{\partial \beta}{\partial x} +
\beta \frac{\partial^2 \beta}{\partial x \partial y} +
\frac{\partial^2 \beta}{\partial t \partial y}\right),
\label{et02}
\ee

\be
G_{03} = - \frac{1}{2}
\frac{\partial^2 \beta}{\partial x \partial z} 
- \frac{\beta}{2} 
\left(2\frac{\partial \beta}{\partial z}
\, \frac{\partial \beta}{\partial x} +
\beta \frac{\partial^2 \beta}{\partial x \partial z} +
\frac{\partial^2 \beta}{\partial t \partial z}\right),
\label{et03}
\ee

\be
G_{11} = \Lambda - \frac{3}{4} \left[
\left(\frac{\partial \beta}{\partial y}\right)^2 
+ \left(\frac{\partial \beta}{\partial z}\right)^2
\right], 
\label{et11}
\ee

\be
G_{12} = \frac{1}{2}\left(
2 \frac{\partial \beta}{\partial y} \, 
\frac{\partial \beta}{\partial x} 
+ \beta \frac{\partial^2 \beta}{\partial x \partial y} 
+ \frac{\partial^2 \beta}{\partial t \partial y}\right),
\label{et12}
\ee

\be
G_{13} = \frac{1}{2}\left(
2 \frac{\partial \beta}{\partial z} \, 
\frac{\partial \beta}{\partial x} 
+ \beta \frac{\partial^2 \beta}{\partial x \partial z} 
+ \frac{\partial^2 \beta}{\partial t \partial z}\right),
\label{et13}
\ee

\be
G_{23} = \frac{1}{2} \frac{\partial \beta}{\partial z} 
\, \frac{\partial \beta}{\partial y},
\label{et23}
\ee

\be
G_{22} = - \Lambda - \frac{\partial}{\partial x}\left[
\frac{\partial \beta}{\partial t}
+ \frac{1}{2} \frac{\partial}{\partial x} (\beta^2)
\right]
- \frac{1}{4}\left[
\left(\frac{\partial \beta}{\partial y}\right)^2
- \left(\frac{\partial \beta}{\partial z}\right)^2
\right],
\label{et22}
\ee

\be
G_{33} = - \Lambda - \frac{\partial}{\partial x}\left[
\frac{\partial \beta}{\partial t}
+ \frac{1}{2} \frac{\partial}{\partial x} (\beta^2)
\right]
+ \frac{1}{4}\left[
\left(\frac{\partial \beta}{\partial y}\right)^2
- \left(\frac{\partial \beta}{\partial z}\right)^2
\right]\,\,.
\label{et33}
\ee

After a long algebraic manipulation of these expressions and
considering the Einstein equations $G_{\mu\nu} = 8 \pi T_{\mu\nu}$,
we obtain the following set of partial differential equations
\be
\frac{4}{3} \Lambda = 8 \pi \left[T_{00} + 2 \beta T_{01} 
+ \left(\beta^2 - \frac{1}{3} \right)T_{11}\right]\,,	
\label{eqset1}
\ee

\be
\frac{1}{2}\frac{\partial^2 \beta}{\partial y^2} + 
\frac{1}{2}\frac{\partial^2 \beta}{\partial z^2} = 
8 \pi (T_{01} + \beta T_{11})\,,
\label{eqset2}
\ee

\be
- \frac{3}{4}\left(\frac{\partial \beta}{\partial y}\right)^2 
- \frac{3}{4} \left(\frac{\partial \beta}{\partial z}\right)^2 
- \Lambda
=  8 \pi T_{11}\,,
\label{eqset3}
\ee

\be
\frac{1}{2}\left(\frac{\partial \beta}{\partial y}\right)^2 
- \frac{1}{2} \left(\frac{\partial \beta}{\partial z}\right)^2 
= 8 \pi (T_{22} - T_{33})\,,
\label{eqset4}
\ee

\be
- \frac{\partial}{\partial x} \left(\frac{\partial \beta}{\partial t}
+ \frac{1}{2} \frac{\partial}{\partial x} (\beta^2)\right) 
- 2 \Lambda
= 8 \pi(T_{33} + T_{22})\,,
\label{eqset5}
\ee

\be
- \frac{1}{2}\frac{\partial^2 \beta}{\partial x \partial y} = 
8\pi (T_{02} + \beta T_{12})\,,
\label{eqset6}
\ee

\be
- \frac{1}{2}\frac{\partial^2 \beta}{\partial x \partial z} = 
8 \pi (T_{03} + \beta T_{13})\,,
\label{eqset7}
\ee

\be
\frac{1}{2}\frac{\partial \beta}{\partial y}\frac{\partial \beta}{\partial z}
= 8 \pi T_{23}\,\,,
\label{eqset8}
\ee

\be
 - \frac{1}{4} \left(\frac{\partial\beta}{\partial y}\right)^2 
- \frac{1}{4} \left(\frac{\partial\beta}{\partial z}\right)^2 
+ \Lambda = 8\pi\left(T_{00} + 2 \beta T_{01} + \beta^2 T_{11}\right)\,\,.
\label{eqset9}
\ee
Note that the set of partial differential equations from Eq.\ \eqref{eqset1}
to Eq.\,\eqref{eqset9} are quite cumbersome to solve if we consider the 
electromagnetic components that appear in the EEMT components $T_{\mu\nu}$,
even if we consider them as constants components.

%%%%%%%%%%%%%%%%%%%%%%%%%%%%%%%%%%%%%%%%%%%%%%%%%%%%%%%%%%%%%%%%%%%%%

\section{Energy conditions} \label{engconds}
\renewcommand{\theequation}{4.\arabic{equation}}
\setcounter{equation}{0}

In this section we will calculate the energy conditions for the
 EEMT and we will see that the inequalities written 
above will be satisfied.  We will find connections between the
components of the electromagnetic field, 
the radiant matter density $\mu$ and the shift vector $\beta$.

%%%%%%%%%%%%%%%%%%%%%%%%%%%%%%%%%%%%%%%%%%%%%%%%%%%%%%%%%%%%%%%%%%%%%

\subsection{Weak Energy Conditions} 

For the weak energy condition, the EMT at each point 
of the spacetime must satisfy the condition
\be
T_{\alpha \sigma} \, u^\alpha u^\sigma \geq 0\,\,,    
\label{weccond}
\ee
where, for any timelike vector $\textbf{u} \, (u_\alpha u^\alpha < 0)$, and 
any zero vector $\textbf{k} \, (k_\alpha k^\alpha = 0)$, for an
observer with unit tangent vector $\textbf{v}$ at a certain point of
the spacetime, the local energy density measured by any observer is
non-negative \cite{HawkingEllis1973}. For the EEMT, the expression 
$T_{\alpha \sigma}\,u^\alpha u^\sigma$ is given by 
\be
T_{\alpha \sigma} \, u^\alpha u^\sigma =  
\frac{1}{2}\Big(E_2^2 + E_3^2\Big) \beta^2 -
\Big(B_3 E_2 - B_2 E_3\Big) \beta +
\frac{1}{2} \Big(B^2 + E^2\Big) + \mu\,\,.
\label{wec}
\ee

Notice that Eq.\ \eqref{wec} is a quadratic function of $\beta$. 
So, to be solved the discriminant of this quadratic equation must
be positive, namely
\be
\Big(B_3 E_2 - B_2 E_3\Big)^2 - 2\Big(E_2^2 + E_3^2\Big)
\left[\frac{1}{2} \Big(B^2 + E^2\Big) + \mu\right]\:\: > \:\:0\,\,,
\label{wec2}
\ee
and we have for the matter density $\mu$ that, 
\be
0\; <\; \mu \:\: < \:\: \frac{\left(B_3 E_2 - B_2 E_3\right)^2}
{2\left(E_2^2 + E_3^2\right)} 
- \frac{1}{2}\Big(B^2 + E^2\Big)\,\,,
\label{wec3}
\ee
which shows a condition between the electromagnetic components,
$$\Big(B_3 E_2 - B_2 E_3\Big)^2\;
> \;\Big(B^2 + E^2\Big)\Big(E^2\,-\,E_1\Big)$$
and that the matter density must have a positive inferior 
minimum value, since the r.h.s.\ of the inequality in Eq.\
\eqref{wec3} must be always positive. This result tells us that
the weak energy condition must be satisfied if we consider both
positive and negative matter density. Solving exactly the inequality
in Eq. \eqref{wec} we have that
\be
\beta_- \;<\;\beta\;<\:\beta_+
\ee
where
\be
\label{wec-sol}
\beta_\pm\,=\, \frac{B_3 E_2 - B_2 E_3}{E^2\,-\,E_1^2}\,\pm\,
\sqrt{\Bigg[\frac{B_3 E_2 - B_2 E_3}{E^2\,-\,E_1^2}\Bigg]^2\,-\,
\frac{B^2+E^2+2\mu}{E^2-E^2_1}}
\ee
and considering a ``weak" warp drive where $\beta^2 \approx 0$ in
Eq.\ \eqref{wec}, then for the weak energy condition to be valid it
requires that
\ba
\beta \:\: > \:\: 
\frac{1}{B_3 E_2 - B_2 E_3}\Bigg[\; \frac 12 \Big(B^2 + E^2\Big) \,+\,\mu
\;\Bigg] \,\,, \\
\mbox{} \nonumber
\label{wec4}
\ea
which tells us that in this case, the shift vector has a limiting 
minimum value condition. This is an interesting result since it implies 
that even for lower order of $\beta$ the warp bubble speed is not 
limited by the weak energy condition. Notice that if $B_3 E_2 \rightarrow 
B_2 E_3$, then the right hand side of this last equation assumes unlimited
values, which could be observed in fact in low strength electromagnetic fields 
and low density of radiant matter. 

%%%%%%%%%%%%%%%%%%%%%%%%%%%%%%%%%%%%%%%%%%%%%%%%%%%%%%%%%%%%%%%%%%%%%

\subsection{Dominant Energy Conditions}

For every timelike vector $u_a$ the following inequalities must be
satisfied
\be
T^{\alpha \beta} \, u_\alpha u_\beta \:\:\geq \:\:0, \qquad \text{and} \qquad 
F^\alpha  F_\alpha \:\: 
\leq \:\: 0\,\,, 
\ee
where $F^\alpha = T^{\alpha \beta} u_\beta$ is a non-spacelike 
vector. We can realize that these conditions mean that for any 
observer, the local energy density appears to be non-negative and the 
local energy flow vector is non-spacelike. In any orthonormal basis 
the energy dominates the other components of the EMT,
\be
T^{00}\:\: \geq \:\:|T^{ab}|\,\,, \ \text{for each} \:\:\ a, b.
\ee
Evaluating the first condition  $T^{\alpha \beta} \, u_\alpha 
u_\beta \geq 0$ for the dominant energy condition gives us the same
result as the weak energy condition seen in Eqs.\ \eqref{wec2} 
and \eqref{wec3}, so this energy condition term can be satisfied
by both positive and negative radiant matter density. Calculating the 
second condition for the dominant energy condition $F^\alpha  
F_\alpha$, it will be given by a fourth degree polynomial on $\beta$
given by the following expression,
\be
F^\alpha  F_\alpha = \omega_{(4)} \beta^4 + \omega_{(3)} \beta^3 + 
\omega_{(2)} \beta^2 + \omega_{(1)} \beta + \omega_{(0)}\,\,,
\label{dec2}
\ee
where $\omega_{(0)},\ldots, \omega_{(4)}$ are implicit functions of the 
spacetime coordinates $(t,x,y,z)$ and also explicit functions in 
terms of the electromagnetic field components. Notice that the 
subindexes of $\omega_{(k)}$'s are not tensor indexes and the terms 
$\omega_{(k)} \beta^k$ are not tensor contractions. The coefficients 
are given by
\be
\omega_{(4)} = \frac{1}{4}\left(E_2^2 + E_3^2\right)^2\,\,,
\ee

\be 
\omega_{(3)} = (E_2^2 + E_3^2)(E_2 B_3 - E_3 B_2)\,\,,
\ee

\be
\omega_{(2)} = \frac{1}{2}(E_2^2 + E_3^2)(E^2 - B^2 - 2 \mu)
- (B_2 E_3 - B_3 E_2)^2\,\,,
\ee

\be 
\omega_{(1)} = (E_2 B_3 - E_3 B_2)(B^2 - E^2 + 2 \mu) \,\,,
\ee
\begin{align}
\nonumber \omega_{(0)} &= - \frac{1}{4} (B^4 + E^2)
-  (E^2 + B^2) \mu
-  \mu^2
\\
\nonumber &- \frac{1}{2} (B_1 E_1 + B_2 E_2 + B_3 E_3)^2
+ \frac{1}{2} (B_1 E_2 - B_2 E_1)^2
\\
&+ \frac{1}{2} (B_1 E_3 - B_3 E_1)^2 
+ \frac{1}{2} (B_3^2 E_2^2 - B_2 E_3)^2\,\,.
\end{align}
Since $F^\alpha  F_\alpha$ is a fourth order polynomial on $\beta$  
with complicated expressions for its components it is a rather 
challenging work to calculate all the roots for this polynomial 
and to find the general requirements for the dominant energy 
condition to be true.   However, a simple and pragmatic way to
impose the validity for this energy condition is to require that
all the coefficients $\omega_{(k)}$ to be a positive value.

%%%%%%%%%%%%%%%%%%%%%%%%%%%%%%%%%%%%%%%%%%%%%%%%%%%%%%%%%%%%%%%%%%%%%

\subsection{Strong Energy Conditions}

The strong energy condition is given by
\be
\left(T_{\alpha \beta} - \frac{1}{2}T \, g_{\alpha \beta} \right) 
u^\alpha u^\beta \:\: \geq \:\: 0 \,\,,
\label{sec}
\ee
for any timelike vector $u$. The expression from the strong energy 
condition in Eq.\,\eqref{sec} for the EEMT and the warp drive metric 
is given by the following equation
\be
 \left(E_2^2 + E_3^2\right) \beta^2 
- 2 \left(B_3 E_2 - B_2 E_3\right) \beta + B^2 + E^2 + \mu \:\: \geq \:\: 0 \,\,,
\label{sec2}
\ee
which is a quadratic $\beta$-inequality. For the strong energy 
condition to be satisfied it is necessary that the discriminant 
of the above quadratic equation be real, that is
\be
\left(B_3 E_2 - B_2 E_3\right)^2 - \left(E_2^2 + E_3^2\right)
\Big[B^2 + E^2 + \mu\Big] \:\: < \:\: 0\,\,,
\ee
which is very similar to the equation we found for the weak
and first dominant energy conditions with the difference of a 
multiplication by a factor of 2
\be
\mu \:\: \leq \:\: \frac{\left(B_3 E_2 - B_2 E_3\right)^2}
{\left(E_2^2 + E_3^2\right)} - \left(B^2 + E^2\right)\,\,,
\label{sec3}
\ee
which has the same meaning of the matter density to satisfy the 
strong energy condition as in Eq. \eqref{wec3} for the weak energy 
condition. The matter density has an inferior bound and it may 
be negative. But the strong energy condition could still be 
satisfied as long as $\mu$ is less than or equal to the right hand side of 
the inequality in Eq.\,\eqref{sec3}.

%%%%%%%%%%%%%%%%%%%%%%%%%%%%%%%%%%%%%%%%%%%%%%%%%%%%%%%%%%%%%%%%%%%%%

\subsection{Null Energy Conditions}

The null energy conditions are satisfied for the null vector 
$\textbf{k}$ since the following inequality must be satisfied
\be
T_{\alpha \sigma} \, k^\alpha k^\sigma \geq 0, \ \text{for any null 
vector} \ k^\alpha.
\ee
Assuming the following vector $k^\alpha = (a,b,0,0)$ where $a$ 
and $b$ can be determined by solving the equation $k_\alpha 
k^\alpha = 0$.   Hence, we can find two possible results
connecting $a$ and $b$,
\be
a = \frac{b}{\beta + 1} \ \qquad \text{or} \qquad \ a = \frac{b}
{\beta - 1}\,\,.
\label{abnullvecsolved}
\ee
Considering this last result, the calculation of $T_{\alpha \sigma}
\, k^\alpha k^\sigma$ is given by
\be
T_{\alpha \sigma} \, k^\alpha k^\sigma =
T_{00} k^0 k^0 + 2 T_{01} k^0 k^1 + T_{11} k^1 k^1\,\,.
\label{nullcondcalc}
\ee
Substituting  Eq.\ \eqref{abnullvecsolved} and that  
$k^0 = a$ and $k^1 = b$ into Eq. \eqref{nullcondcalc}, 
the result of the null energy condition for the EEMT is a 
complicated function of $\beta$ given by
\be
T_{\alpha \sigma} \, k^\alpha k^\sigma = 
\omega_{(4)} \beta^4 + \omega_{(3)} \beta^3 + \omega_{(2)} \beta^2 
+ \omega_{(1)} \beta + \omega_{(0)}\,\,,
\label{nec}
\ee
where the coefficients $\omega_{(k)}$ are also functions of $\beta$ 
given by the expressions
\be
\omega_{(4)} = \frac{a^2}{2}(E_2^2 + E_3^2)\,\,,
\ee

\be
\omega_{(3)} = a^2 (B_2 E_3 - B_3 E_2) - a b (E_2^2 + E_3^2)\,\,,
\ee

\be
\omega_{(2)} = \frac{1}{2}\left[a^2 (B^2 - 2 B_1^2 - E_1^2)
+ 4 a b (B_3 E_2 - B_2 E_3)
+ (b^2 - 2 a^2) (E_2^2 + E_3^2)\right] \,\,,
\ee

\be
\omega_{(1)} = a b (B_1^2 - B_2^2 - B_3^2 + E^2) 
+ (a^2 - b^2) (B_3 E_2 - B_2 E_3)\,\,,
\ee

\be
\omega_{(0)} = \frac{a^2}{2}(B^2 + E^2)
+ \frac{b^2}{2}(B^2 - 2B_1^2 + E^2 - 2 E_1^2)
+ 2 a b (B_2 E_3 - B_3 E_2) + a^2 \mu \,\,,
\ee
which are hard-working algebraic expressions concerning the null 
energy condition.   The next step would be to solve them 
in a general way and to find the specific connections for 
the energy  condition to be valid.

%%%%%%%%%%%%%%%%%%%%%%%%%%%%%%%%%%%%%%%%%%%%%%%%%%%%%%%%%%%%%%%%%%%%%

\section{Another expression of the energy momentum tensor}
\renewcommand{\theequation}{5.\arabic{equation}}
\setcounter{equation}{0} 

In this section we will introduce a specific observer that simplifies 
the EMT components. We will use this approach as a 
laboratory relative to the warp drive in an electromagnetic field 
background before analyzing and solving the generalized Einstein 
equations.

%%%%%%%%%%%%%%%%%%%%%%%%%%%%%%%%%%%%%%%%%%%%%%%%%%%%%%%%%%%%%%%%%%%%%

\subsection{Choosing a specific observer}

Let us consider that the electric and magnetic fields are orthogonal 
to each other, like in an electromagnetic wave. In addition, they do
not obey the wave equation, but they do satisfy the relation below,
\begin{equation}
\label{plane-wave}
E_i B^i = 0.
\end{equation}
Now, let us assume that from the point of view of this observer the 
electric field has  only one component, $E_1$, which is the only one 
non zero component and it points to the direction of the particle 
trajectory inside the warp bubble. Hence, from Eq.\ \eqref{plane-wave}, 
it is straightforward that $B_1 = 0$, but we can still have the other 
magnetic components. Having said that, the non zero components of the
EEMT can  be computed such that,  
\be
4\pi T_{00} = \mu + \frac{1}{2}(B^2 + E_1^2)
+ \frac{\beta^2}{2}\left(B^2 - E_1^2\right)\,\,,
\label{simpeemt00}
\ee

\be
4\pi T_{01} = \frac{\beta}{2}\left(E_1^2 - B^2\right)\,\,,
\label{simpeemt01}
\ee

\be
4\pi T_{02} = B_3 E_1\,\,, 
\label{simpeemt02}
\ee

\be
4 \pi T_{03} = - B_2 E_1\,\,,
\label{simpeemt03}
\ee

\be
4\pi T_{11} = \frac{1}{2}\left(B^2 - E_1^2\right)\,\,, 
\label{simpeemt11}
\ee

\be
4\pi T_{22} = \frac{1}{2} \left(B_3^2 - B_2^2 + E_1^2\right)\,\,,
\label{simpeemt22} 
\ee

\be
4\pi T_{23} = - B_2 B_3 \,\,,
\label{simpeemt23} 
\ee

\be
4\pi T_{33} =  \frac{1}{2}\left(B_2^2 - B_3^2 + E_1^2\right)\,\,.
\label{simpeemt33}
\ee

Notice that $B^2 = B_2^2 + B_3^2$ and now that $E = E_1$. 

%%%%%%%%%%%%%%%%%%%%%%%%%%%%%%%%%%%%%%%%%%%%%%%%%%%%%%%%%%%%%%%%%%%%%

\subsection{Energy conditions for the specific electromagnetic field}

Now we will use our specific choice of the electromagnetic field, namely, 
$\mathbf{E} = (E_1,0,0)$ and $\mathbf{B} = (0, B_2, B_3)$ in the energy 
conditions. We have demonstrated in the last section that both the weak and 
strong energy conditions could be satisfied in a general manner, 
but the dominant and null conditions require a hard-working 
calculation to show that they are valid.

%%%%%%%%%%%%%%%%%%%%%%%%%%%%%%%%%%%%%%%%%%%%%%%%%%%%%%%%%%%%%%%%%%%%%

\subsubsection{Weak energy condition for the simplified choice of EEMT}
 
Using the results obtained in the last section  together with 
the specific choice of the electromagnetic field, we can 
find that the weak energy condition from Eq. \eqref{wec} can be 
satisfied if 
\be
T_{\alpha \sigma} \, u^\alpha u^\sigma =  
\frac{1}{2} \left(B^2 + E^2\right) + \mu \:\: \geq \:\: 0\,\,,
\label{simpwec}
\ee
which means that even with no radiant matter density, i.e., $\mu = 
0$, the weak energy inequality is still non-negative. Besides, even with 
a negative matter density, it could still be positive for this energy
condition to be satisfied if the electromagnetic field strength 
$\left(B^2 + E_1^2\right)/2$ is bigger than the matter 
density $\mu$ as in Eq. \eqref{simpwec}.

%%%%%%%%%%%%%%%%%%%%%%%%%%%%%%%%%%%%%%%%%%%%%%%%%%%%%%%%%%%%%%%%%%%%%

\subsubsection{Strong energy condition for the simplified choice of EEMT}

From Eq. \eqref{sec2} the inequality simplifies to
\be
\left(T_{\alpha \beta} - \frac{1}{2}T \, g_{\alpha \beta} \right) 
u^\alpha u^\beta \:=\: B^2 + E^2 + \mu \:\: \geq \:\: 0\,\,,
\label{simpsec2}
\ee
and it is clear that, since $B^2 + E^2$ is obviously always positive, 
the strong energy condition is valid for the simplified choice of
electromagnetic field  since the matter density $\mu$ is positive. In the
case of a negative matter density, the modulo of electromagnetic
energy density $B^2 + E^2$ must be larger than the negative
matter density $\mu$. 

%%%%%%%%%%%%%%%%%%%%%%%%%%%%%%%%%%%%%%%%%%%%%%%%%%%%%%%%%%%%%%%%%%%%%

\subsubsection{Dominant energy condition for the simplified choice of EEMT}

For the dominant energy inequality, the first condition 
$T^{\alpha \beta} \, u_\alpha u_\beta \geq 0$, still gives us 
the same result as the one from the weak energy inequality.   But now with
this simplified choice for the electromagnetic field, the second 
requirement for the dominant energy condition is given by
\be
F^\alpha  F_\alpha = -\mu^2 - (E^2 + B^2)\mu 
- \frac{1}{2} (E^2 - B^2)^2 - \frac{1}{2} E^2 B^2\,\,,
\label{simpdec2}
\ee
which is a quadratic function of the matter density $\mu$. The sign of this
expression will depend on the value of the electromagnetic components, of
course. But, for real solutions, we must have that $E^2B^2 \geq
(E^2-B^2)^2/2$.

%%%%%%%%%%%%%%%%%%%%%%%%%%%%%%%%%%%%%%%%%%%%%%%%%%%%%%%%%%%%%%%%%%%%%

\subsubsection{Null energy condition for the simplified choice of EEMT}

Recovering the null energy condition from Eq.\,\eqref{nec}, we have 
that
\be
T_{\alpha \sigma} \, k^\alpha k^\sigma =
\frac{a^2}{2}(B^2 - E^2)\beta^2 
+ a b (E^2 - B^2)\beta + 
\frac{1}{2}(a^2 + b^2)B^2 + \frac{1}{2}(a^2 - b^2)E^2  + a^2 \mu\,\,,
\label{simpnec}
\ee

Disregarding the fact that $a$ and $b$ are functions of $\beta$, we
will use the assumption that Eq. \eqref{simpnec} is a quadratic 
function of $\beta$ and we will impose that the discriminant of this equation 
must be positive or zero for real solutions.   Hence, the null energy
condition might be satisfied for any value of $\beta$ if
\be
a^4 (E^2 - B^2)(\mu + B^2 + E^2) \geq 0\,\,.
\label{simpnec2}
\ee
which means that, considering a positive $\mu$, the main condition is
$E^2 \geq B^2$.

Table \ref{table:nonlin} summarizes all the necessary requirements for 
the energy conditions to be valid simultaneously for the specific
simplified choice of the electromagnetic field.
\begin{table}[ht]
\caption{Summary for the energy conditions} 
\centering 
\begin{tabular}[c]{l @{\hspace{50pt}} l} 
\hline\hline 
Energy condition & Results \\ [0.5ex] 
\hline 
Weak         &  $\frac{1}{2} \left(B^2 + E^2\right) + \mu \geq 0$  \\ 
Strong       &  $B^2 + E^2 + \mu \geq 0$  \\
Dominant     &  $ E^2B^2\geq   (E^2-B^2)^2/2$ \\
Null         &  $ E^2 \geq B^2\,, \mbox{for}\; \mu>0.$ \\ 
[1ex] 
\hline 
\end{tabular}
\label{table:nonlin} 
\end{table}

%%%%%%%%%%%%%%%%%%%%%%%%%%%%%%%%%%%%%%%%%%%%%%%%%%%%%%%%%%%%%%%%%%%%%

\subsection{Solving the Einstein equations}

For the simplified choice of the electromagnetic field, the set of partial
differential Eqs. \eqref{eqset1} to \eqref{eqset9}, that resulted
from algebraic simplifications of the Einstein equations are
\be
\frac{4}{3} \Lambda = 8 \pi \left(\mu + \frac{1}{3} B^2 + 
\frac{2}{3}E_1^2\right)\,,	
\label{eqsetsimp1}
\ee

\be
\frac{1}{2}\frac{\partial^2 \beta}{\partial y^2} + 
\frac{1}{2}\frac{\partial^2 \beta}{\partial z^2} = 0\,,
\label{eqsetsimp2}
\ee

\be
- \frac{3}{4}\left(\frac{\partial \beta}{\partial y}\right)^2 
- \frac{3}{4} \left(\frac{\partial \beta}{\partial z}\right)^2 
- \Lambda =  4 \pi \left(B^2 - E_1^2 \right) \,,
\label{eqsetsimp3}
\ee

\be
\frac{1}{2}\left(\frac{\partial \beta}{\partial y}\right)^2 
- \frac{1}{2} \left(\frac{\partial \beta}{\partial z}\right)^2 
= 8 \pi \left(B_3^2 - B_2^2\right)\,,
\label{eqsetsimp4}
\ee

\be
- \frac{\partial}{\partial x} \left(\frac{\partial \beta}{\partial t}
+ \frac{1}{2} \frac{\partial}{\partial x} (\beta^2)\right) 
- 2 \Lambda = 8 \pi E_1^2\,,
\label{eqsetsimp5}
\ee

\be
- \frac{1}{2}\frac{\partial^2 \beta}{\partial x \partial y} = 
8 \pi B_3 E_1\,,
\label{eqsetsimp6}
\ee

\be
- \frac{1}{2}\frac{\partial^2 \beta}{\partial x \partial z} = 
- 8 \pi B_3 E_1\,,
\label{eqsetsimp7}
\ee

\be
\frac{1}{2}\frac{\partial \beta}{\partial y}\frac{\partial \beta}{\partial z}
= - 8 \pi B_2 B_3\,\,,
\label{eqsetsimp8}
\ee

\be
 - \frac{1}{4} \left(\frac{\partial\beta}{\partial y}\right)^2 
- \frac{1}{4} \left(\frac{\partial\beta}{\partial z}\right)^2 
+ \Lambda = 4\pi\left(2\mu + E_1^2 + B^2\right)\,\,,
\label{eqsetsimp9}
\ee

%%%%%%%%%%%%%%%%%%%%%%%%%%%%%%%%%%%%%%%%%%%%%%%%%%%%%%%%%%%%%%%%%%%%%%
\subsection{Divergence of the specific electromagnetic field}

In this section we will calculate the divergence of the EEMT 
considering that the components of the electromagnetic field are
functions of the spacetime coordinates $x^\mu$. The results are as
follows,
\be
{T^{0 \alpha}}_{;\alpha}  = - \frac{\partial (\beta B^2)}{\partial x} 
- \frac{1}{2} \frac{\partial (B^2 + E_1^2)}{\partial t} 
- \frac{\partial (B_2 E_1)}{\partial z} 
+ \frac{\partial (B_3 E_1)}{\partial y} 
- \frac{\partial (\mu \beta)}{\partial x} 
- \frac{\partial\mu}{\partial t}  \,\,,
\label{divcomp0}
\ee

\be
{T^{1 \alpha}}_{;\alpha} =  \frac{1}{2}
\frac{\partial (B^2 - E_1^2)}{\partial x}  \,\,,
\label{divcomp1}
\ee

\be
{T^{2 \alpha}}_{;\alpha} = -  \frac{\partial (\beta B_3 E_1)}{\partial x} 
- \frac{\partial (B_2 B_3)}{\partial z} 
- \frac{\partial (B_3 E_1)}{\partial t} 
+ \frac{1}{2}\frac{\partial}{\partial y}(B_3^2 - B_2^2 + E_1^2) \,\,,
\label{divcomp2} 
\ee

\be
{T^{3 \alpha}}_{;\alpha} = \frac{\partial (\beta B_2 E_1)}{\partial x} 
+ \frac{\partial (B_2 E_1)}{\partial t} 
- \frac{\partial (B_2 B_3)}{\partial y} 
+ \frac{1}{2} \frac{\partial}{\partial z} (B_2^2 - B_3^2 + E_1^2)\,\,.
\label{divcomp3}  
\ee

Considering that $E_1$, $B_1$ and $B_2$ are constants, Eqs.\ 
\eqref{divcomp0} to \eqref{divcomp3} can be written such as
\be
{T^{0 \alpha}}_{;\alpha}  = - B^2 \frac{\partial \beta}{\partial x} 
- \frac{\partial (\mu \beta)}{\partial x} 
- \frac{\partial\mu}{\partial t} 
\label{divcompsimp0}
\ee

\be
{T^{1 \alpha}}_{;\alpha} =  0 
\label{divcompsimp1}
\ee

\be
{T^{2 \alpha}}_{;\alpha} = - B_3 E_1 \frac{\partial \beta}{\partial x} 
\label{divcompsimp2} 
\ee

\be
{T^{3 \alpha}}_{;\alpha} = B_2 E_1 \frac{\partial \beta}{\partial x} 
\label{divcompsimp3}  
\ee
Imposing that the EEMT must be conserved, i.e., the divergence must
be zero, which shows that Eq.\,\eqref{divcompsimp0} is a continuity
equation, but Eq.\ \eqref{divcompsimp2} implies that either $B_3 = 0$,
or $E_1 = 0$ or $\frac{\partial \beta} {\partial x} = 0$, and Eq.\
\eqref{divcompsimp3} implies that either $B_2 = 0$, or $E_1 = 0$ or
$\frac{\partial \beta}{\partial x} = 0$. Next we will analyze how each
one of these cases can affect the set of Eqs.\ \eqref{eqsetsimp1} to
\eqref{eqsetsimp9}.

%%%%%%%%%%%%%%%%%%%%%%%%%%%%%%%%%%%%%%%%%%%%%%%%%%%%%%%%%%%%%%%%%%%%%

\subsubsection{Case $\frac{\partial \beta}{\partial x} = 0$ and 
$E_1 = 0$}

These conditions are satisfied simultaneously and it can be seen 
from Eq. \eqref{eqsetsimp5}.  Moreover, Eqs. \eqref{eqsetsimp6} 
and \eqref{eqsetsimp7} are identically zero and the set of Einstein 
differential equations from Eq. \eqref{eqsetsimp1} to 
\eqref{eqsetsimp9} can be written such as

\be
\Lambda = 6 \pi \mu + 2 \pi B^2\,,	
\label{eqsetbeta1}
\ee

\be
\frac{\partial^2 \beta}{\partial y^2} + 
\frac{\partial^2 \beta}{\partial z^2} = 0\,,
\label{eqsetbeta2}
\ee

\be
\left(\frac{\partial \beta}{\partial y}\right)^2 
+ \left(\frac{\partial \beta}{\partial z}\right)^2 
 = - \frac{4}{3}\Lambda  - \frac{16\pi}{3} B^2 \,,
\label{eqsetbeta3}
\ee

\be
\left(\frac{\partial \beta}{\partial y}\right)^2 
- \left(\frac{\partial \beta}{\partial z}\right)^2 
= 16 \pi \left(B_3^2 - B_2^2\right)\,,
\label{eqsetbeta4}
\ee

\be
\frac{\partial \beta}{\partial y}\frac{\partial \beta}{\partial z} 
= - 16 \pi B_2 B_3\,\,,
\label{eqsetbeta5}
\ee

\be
\left(\frac{\partial\beta}{\partial y}\right)^2 
+ \left(\frac{\partial\beta}{\partial z}\right)^2 
= - 4 \Lambda - 16 \pi\left(2\mu + B^2\right)\,\,.
\label{eqsetbeta6}
\ee

The above set of equations imply that the cosmological constant is
null and the following relation between the electromagnetic field and 
the matter density
\be
\mu = - \frac{1}{3} B^2.
\label{matterdenneg}
\ee
So, the matter density will always be negative for this case, but the
energy conditions will still be satisfied. The shift vector will not 
depend on the $x$ spacetime coordinate, and there is no Burgers 
equation, i.e., no shock wave, but $\beta$ is function of $(t,y,z)$
and it also satisfies the Laplace equation according to Eq.\,\eqref{eqsetbeta2}.
Both Eqs. \eqref{eqsetbeta3} and \eqref{eqsetbeta6} are specific cases of the
well known Eikonal equation, and they imply that the solution for the
shift vector is not unique and may have a complex component. For this
case the EMT in matrix form is given by, 
\be
T_{\alpha \nu} =
\begin{pmatrix}
-\frac{1}{2}\mu(1 + \beta^2) & -\frac{1}{2}\beta B^2 & 0              & 0  \\
-\frac{1}{2}\beta B^2       & -\frac{1}{2}B^2       & 0              & 0 \\
0                           & 0                     & \frac{1}{2}
	(B_3^2 - B_2^2) & - B_2 B_3 \\
0                           & 0                     & - B_2 B_3     
	& - \frac{1}{2}(B_3^2 - B_2^2)
\end{pmatrix}\,\,.
\label{emmtsimp1}
\ee

It is clear that since the matter density is negative, i.e., $\mu < 0$, 
then the energy density for the energy momentum tensor is positive, namely,  
$T_{00} > 0$.

%%%%%%%%%%%%%%%%%%%%%%%%%%%%%%%%%%%%%%%%%%%%%%%%%%%%%%%%%%%%%%%%%%%%%

\subsubsection{Case $B_3 = 0$ and $B_2 = 0$}

For this case it is straightforward to see that $B_2 = B_3 = 0$ 
implies that $\frac{\partial \beta}{\partial y} = 0$ and 
$\frac{\partial \beta}{\partial z} = 0$ from Eqs. 
\eqref{eqsetsimp6} to \eqref{eqsetsimp8}. The
set of equations for this case is the following

\be
\Lambda = 4 \pi E_1^2 \,,	
\label{eqsetmag1}
\ee

\be
\mu = 0 \,,	
\label{eqsetmag2}
\ee

\be
- \frac{\partial}{\partial x} \left(\frac{\partial \beta}{\partial t}
+ \frac{1}{2} \frac{\partial}{\partial x} (\beta^2)\right) = 
2 \Lambda + 8 \pi E_1^2\,.
\label{eqsetmag3}
\ee

This result violates both null and dominant energy conditions, but it
is an interesting theoretical result since both matter density $\mu$ 
and magnetic field \textbf{B} are null, but the cosmological constant 
is positive and proportional to the electric field energy as seen 
in Eq.\ \eqref{eqsetmag1}.
\be
T_{\mu \nu} =
\begin{pmatrix}
-\frac{1}{2}E^2(1 - \beta^2) & -\frac{1}{2}\beta E^2 & 0              & 0  \\
-\frac{1}{2}\beta E^2       & -\frac{1}{2}E^2       & 0              & 0 \\
0                           & 0                     & \frac{1}{2}E^2 & 0 \\
0                           & 0                     & 0              & \frac{1}{2}E^2
\end{pmatrix}.
\label{emmtsimp2}
\ee
which is the final and symmetric form of the EEMT.  Notice the presence
of  only the electric field and the shift function since $B=0$.

%%%%%%%%%%%%%%%%%%%%%%%%%%%%%%%%%%%%%%%%%%%%%%%%%%%%%%%%%%%%%%%%%%%%%

\section{Conclusions and final remarks}
\renewcommand{\theequation}{6.\arabic{equation}}
\setcounter{equation}{0}

In this work we have investigated the solutions of the Einstein 
equations for the Alcubierre warp drive metric with the choice of 
dust and electromagnetic field energy-momentum tensor (EMT) as 
possible sources of global superluminal particle speeds, that is, 
warp velocities. The Einstein equations were analyzed and all
results concerning the components of the electromagnetic field were
discussed. The energy conditions were presented as functions of the
electromagnetic field components, and we have established conditions
on these components such that they obey, or not, the energy conditions 
the warp drive metric. The connections found between electromagnetic
field and the superluminal effects of the warp drive resulting from
the Einstein field equations, namely, being solutions of them, are
new. We have also discussed the potential dynamic consequences of
these results on the warp bubble dynamics.

We found that the energy conditions can be satisfied for positive and 
negative matter density, requirements which were summarized in a
table. We also showed that the null divergence of the electromagnetic
tensor results in some interesting cases, such as, the matter density
becoming negative and proportional to the magnetic field energy density,
the shift vector $\beta$ being a function of only $(t,y,z)$, and the 
absence of the Burgers equation as found in Ref.\ \cite{nosso-1}.
Nevertheless, a specific case of Eikonal equation has to be solved,
leading to a wave equation, in order to find solutions for $\beta$.
In another case we found a violation of both the dominant and null
energy conditions, the matter density and the magnetic field are null
and the cosmological constant is proportional to the electric energy
density. 

\section*{Acknowledgments}

E.M.C.A. thanks CNPq (Conselho Nacional de Desenvolvimento Cient\'ifico
e Tecnol\'ogico), Brazilian federal scientific supporting agency, for
partial financial support, grant number 406894/2018-3.

%%%%%%%%%%%%%%%%%%%%%%%%%% Bibliography %%%%%%%%%%%%%%%%%%%%%%%%%%%%%%


\begin{thebibliography}{}

\bibitem{Alcubierre1994}
M.\ Alcubierre,  \textit{The warp drive: hyper-fast travel 
within general relativity}, Class.\ Quant.\ Grav.\ 11 (1994) L73,  
arXiv: gr-qc/0009013.

\bibitem{FordRoman1996}
L.H.\ Ford and T.A.\ Roman, \textit{Quantum Field Theory Constrains 
Traversable Wormhole Geometries}, Phys.\ Rev.\ D 53 (1996) 5496, 
arXiv: gr-qc/9510071.

\bibitem{Pfenning1997}
M.J.\ Pfenning and L.H.\ Ford \textit{The unphysical nature of 
Warp Drive}, Class. Quant.\ Grav.\ 14 (1997) 1743, 
arXiv: gr-qc/9702026.

\bibitem{Krasnikov1998}
S.V.\ Krasnikov \textit{Hyperfast Interstellar Travel in General 
Relativity}, Phys.\ Rev.\ D 57 (1998) 4760, arXiv: gr-qc/9511068.

\bibitem{EveretRoman1997}
A.\ Everett and T.A.\ Roman, \textit{A Superluminal 
Subway: The Krasnikov Tube}, Phys.\ Rev.\ D 56 (1997) 2100, 
arXiv: gr-qc/9702049. 

\bibitem{Lobo2002}
F.S.N.\ Lobo and P.\ Crawford, \textit{Weak Energy Condition 
Violation and Superluminal Travel}, Lect.\ Notes Phys.\ 617 
(2003) 277, arXiv: gr-qc/0204038.

\bibitem{Lobo2003}
F.S.N.\ Lobo and P.\ Crawford, \textit{Weak Energy Condition 
Violation and Superluminal Travel}. In: Fern\'andez-Jambrina L., 
Gonz\'alez-Romero L.M. (eds), \textit{Current Trends in Relativistic 
Astrophysics}, Springer, Berlin, 2003, arXiv: gr-qc/0204038.

\bibitem{Broeck1999}
C.\ Van Den Broeck,  \textit{A warp drive with more reasonable 
total energy}, Class.\ Quant.\ Grav.\ 16 (1999) 3973, arXiv: gr-qc/9905084. 

\bibitem{Natario2002}
J.\ Natario, \textit{Warp Drive With Zero Expansion},
Class. Quant.\ Grav.\ 19 (2002) 1157, arXiv: gr-qc/0110086.

\bibitem{LoboVisser2004}
F.S.N.\ Lobo and M. Visser, \textit{Linearized warp drive and the 
energy conditions}, arXiv: gr-qc/0412065.

\bibitem{White2003}
H.G.\ White, \textit{A Discussion of Space-Time Metric 
Engineering}, Gen.\ Relat.\ Grav.\ 35 (2003) 2025. 

\bibitem{White2011}
H.G.\ White, \textit{Warp Field Mechanics 101}, 
J.\ Brit.\ Interplanetary Society 66 (2011) 242.

\bibitem{cleaver1}
J.\ Lee and G.\ Cleaver, \textit{Effects of External Radiation on an
Alcubierre Warp Bubble}, Physics Essays 29 (2016) 201

\bibitem{cleaver2}
J.\ Lee and G.\ Cleaver, \textit{The Inability of the White-Juday
Warp Field Interferometer to Spectrally Resolve Spacetime Distortions},
Int. J.\ Modern Phys.: Advances in Theory and Application 2 (2017) 35;
arXiv:1407.7772

\bibitem{cleaver3}
B.\ Mattingly, A.\ Kar, M.\ Gorban, W.\ Julius, C.\ Watson, M.D.\ Ali,
A.\ Baas, C.\ Elmore, J.\ Lee, B.\ Shakerin, E.\ Davis, and G.\ Cleaver,
\textit{Curvature Invariants for the Accelerating Natario Warp Drive},
Particles 3 (2020) 642-659; arXiv:2008.03366

\bibitem{nosso-1}    
O. L. Santos-Pereira, E. M. C. Abreu and M. B. Ribeiro, 
\textit{Dust content solutions for the Alcubierre warp drive 
spacetime}, Eur. Phys. J. C 80 (2020) 786, arXiv: 2008.06560 [gr-qc].

% incluída
\bibitem{nosso-2}    
O. L. Santos-Pereira, E. M. C. Abreu and M. B. Ribeiro, 
\textit{Fluid Dynamics in the Warp Drive Spacetime Geometry},
Eur. Phys. J. C 81 (2021) 133, arXiv: 2101.11467 [gr-qc].

\bibitem{dInverno1992}
R. d'Inverno, \textit{Introducing Einstein's Relativity},
Clarendon Press, UK, 1992.

%\bibitem{Alcubierre2012}
%M.\ Alcubierre, \textit{Introduction to 3+1 Numerical Relativity}, 
%Oxford University Press, UK, 2012.

%\bibitem{DeWitt1979}
%B.S.\ DeWitt,  in \textit{General Relativity: An Einstein Centenary 
%Survey}, S.W.\ Hawking and W.\ Israel (eds), Cambridge University 
%Press, UK, 1980.

\bibitem{HawkingEllis1973}
S.W.\ Hawking and  G.F.R.\ Ellis,\textit{The Large Scale 
Structure of Space-Time}, Cambridge University Press, UK, 1973.

%\bibitem{Evans2010}
%L.C.\ Evans, \textit{Partial Differential Equations},
%Graduate Studies in Mathematics, AMS, 19 (2010) 749.

%\bibitem{turcos} 
%T.\ Ceylan and B.\ Okutmu\c{s}tur,
%\textit{Finite volume approximation of the relativistic Burgers
%equation on a Schwarzschild (anti-)de Sitter spacetime}, Turk. J.\
%Math.\ 41 (2017) 1027; \textit{The relativistic Burgers equation on
%a de Sitter spacetime. Derivation and finite volume approximation},
%Int.\ J.\ Pure.\ Math.\ 2 (2015) 20; P.G.\ LeFloch, H.\ Makhlof
%and B.\ Okutmu\c{s}tur, \textit{Relativistic Burgers equation on
%curved spacetime. Derivation and finite volume approximation},
%arXiv: 1206.3018.

%\bibitem{Forsyth1906}
%A.R.\ Forsyth, \textit{Theory of differential equations},
%Cambridge University Press, UK, 1906. 

%\bibitem{Bateman1915}
%H.\ Bateman. \textit{Some recent researches on the motion of fluids},
%Monthly Weather Review, 43 (1915) 163.

%\bibitem{Burgers1948}
%J.M.\ Burgers, \textit{A mathematical model illustrating the 
%theory of turbulence}, Adv. Appl. Mech. Vol. 1, pp. 
%171-199, Elsevier, 1948.

%\bibitem{Cole1951}
%J.D.\ Cole. \textit{On a quasi-linear parabolic equation occurring
%in aerodynamics}, Quart.\ Appl.\ Math.\ 9 (1951) 225.

\end{thebibliography}
\end{document}